\documentstyle[psfig]{mn2e}

\title[A starquake model for Vela pulsar]{A starquake model for Vela pulsar}

\author[P. S. Negi]{P. S. Negi$^{1}$\thanks{E-mail:
negi@upso.ernet.in; psnegi\_nainital@yahoo.com} \\
$^{1}$ Department of Physics, Kumaun University,
              Nainital - 263 002, India \\
}
\date{Accepted ------ .
      Received ------ ;
      }
\pagerange{\pageref{firstpage}--\pageref{lastpage}} 
\pubyear{2005}

\begin{document}

\maketitle

\label{firstpage}

\begin{abstract}
The measured values of glitch healing parameter, $Q$, of the Vela pulsar
are found to be inconsistent with the starquake mechanism of glitch generation in  various neutron star (NS) models, based upon the parameterized equations of state (EOSs)
of dense nuclear matter. Since such models correspond to an unrealistic mass range $\leq 0.5M_\odot$
for the pulsar, if the observational constraints of the fractional moment of inertia
of the core component ($I_{\rm core}/I_{\rm total} \leq 0.2$) which equals the
glitch healing parameter, $Q$, in the starquake model, are imposed on these models.
However, we show that these observational constraints yield a realistic mass range for NS models,
corresponding to a core given by the stiffest equation of 
state (EOS), $dP/dE = 1$ (in geometrized units) and the envelope is characterized by the well known EOS 
of adiabatic polytrope ${\rm d}lnP/{\rm d}ln\rho = \Gamma_1 $), if the continuity of the $adiabatic \,speed\, of sound\, (v = \sqrt(dP/dE)$) together
with pressure ($P$), energy-density ($E$), and the two metric parameters
($\nu$, and $\lambda$) is assured at the core-envelope boundary of the models and this boundary is worked out
on the basis of the `compatibility criterion' for
hydrostatic equilibrium. The models yield the stable sequence of NS
masses in the range, $1.758 M_\odot \leq M \leq 2.2M_\odot$, corresponding to the values of glitch healing
parameter range, $0 \leq Q \leq 0.197$, for a choice of the `transition density', 
$E_b = 1.342 \times 10^{15}{\rm\,g\,cm}^{-3}$, at the core-envelope boundary. 
The maximum stable value of $2.2M_\odot$
in this sequence, in fact, corresponds to the lowest 
 possible upper bound on NS masses calculated in the literature, on the basis of modern
 EOSs for NS matter. The models yield the surface redshift $z_R \simeq 0.6913$ and mass
 $M \simeq 2.153 M_\odot$ for the ``central'' weighted mean value, $Q = 0.12 \pm 0.07$, of the glitch 
healing parameter of the Vela pulsar. This value of mass can increase slightly upto
$M \simeq 2.196 M_\odot$, whereas the surface redshift can increase upto the value
$z_R \simeq 0.7568$ (which represents an ultra-compact object (UCO; $z_R \geq 0.73$)),
if the observational constraint of the upper weighted mean value of $Q \simeq 0.19$ is imposed on these models. However,
for the lower weighted mean value of $Q \simeq 0.05$, the mass and surface redshift
can decrease upto the values of $M \simeq 2.052 M_\odot$ and $z_R \simeq 0.6066$ respectively.
These results set the lower bound on the 
 energy of a gravitationally redshifted radiation in the rather narrow range of 0.291 - 0.302 MeV. 
 The observation of the lower bound on the energy
 of a $\gamma$-ray pulse at about 0.30 MeV from the Vela pulsar in 1984 is in excellent agreement with this
  result, provided this energy could be interpreted as the 
 energy of a gravitationally redshifted electron-positron annihilation radiation from the star's surface.

\end{abstract}

\begin{keywords}
dense matter - equation of state - stars:
                neutron - stars: pulsars: individual: Vela
\end{keywords}

\section{Introduction}
The glitch (sudden increase in the rotational velocity of a pulsar) data yield the important
information regarding the internal structure
of neutron stars (NSs), since it provide the best tool for
testing various glitch models which are actually based upon the internal structure
of NSs. At present, there are two well accepted glitch models available in the literature: (i) the starquake
(Baym \& Pines 1971; Ruderman 1976; Alpar et al 1996) and (ii) the vortex unpinning (Anderson \& Itoh 1975; Alpar et al 1993) models. And, both
of them, in fact, agreed, in general, upon the same conventional notion that a NS may be 
considered as a two-component structure, a superfluid interior core which contains most of the NS's mass
surrounded by a rigid crust which contains only a few percent of the total mass (here the term `crust'
is used for the solid crust plus other interior part of the star (strongly coupled to it) right upto
the superfluid core; we shall call this portion as `envelope' of the star). The envelope of conventional 
NS models is characterized by different equations of state (EOSs) in an appropriate sequence below
a `fiduciary' transition density, $E_b$, and the
core is sometimes characterized by the extreme causal EOS, $dP/dE = 1$  beyond $E_b$, in order to calculate an
upper bound on NS masses. Each member of the stable mass-radius sequence of such models may represent 
a `realistic' case. 

At present, the most extensive and accurate glitch data for Crab and Vela pulsars
are available in the literature. Crawford \& Demia\'{n}ski (2003) have collected the all measured values of
glitch healing parameter, $Q$, for Crab and Vela pulsars, which is defined in starquake glitch model as the 
fractional moment of inertia, i.e. the ratio of the moment of
inertia of the superfluid core, $I_{\rm core}$, to the moment of inertia of the entire configuration,
$I_{\rm total}$, as 
\begin{equation}
Q = \frac{I_{\rm core}}{I_{\rm total}}.
\end{equation}
They have calculated that 21 measured values
of $Q$ for Crab glitches yields a weighted mean of $Q = 0.72 \pm 0.05$, and the range of $Q 
\geq 0.7$ encompasses the observed distribution for the Crab pulsar. In order to test the starquake model for 
Crab pulsar, they have computed $Q$ (as given by Eq.(1)) values for seven representative EOSs of dense
 nuclear matter, covering a range of neutron star masses. Their 
  study shows that the much larger values of $Q(\geq 0.7)$ for the Crab pulsar is fulfilled by all, but 
  the six EOSs considered in the study corresponding to a `realistic' neutron star mass range 
$1.4\pm 0.2M_\odot$. On the other hand, a weighted mean value of the 11 measurements for Vela yields a much
smaller value of $Q(= 0.12 \pm 0.07)$ and the all estimates for Vela agree with the likely range of $Q \leq 0.2$.
Thus, their results are found to be consistent with the starquake model predictions for the Crab pulsar. They 
have also concluded that the much smaller values of $Q \leq 0.2$ for Vela pulsar are inconsistent with the
 starquake model predictions, since the implied Vela mass based upon their models corresponds to a value $ \leq 0.05M_\odot$, which is too low as compared to the `realistic' NS mass range.
  
However, it seems really surprising that if the NS's internal structure is described by the
same conventional NS models (as mentioned above), why different kinds of glitch mechanisms are required for the explanation 
of a glitch! It is apparent from the study of weighted mean value of the glitch healing parameter, $Q$, for the Vela pulsar that one would require a 
two-component model of NS such that more than eighty percent of the total moment of inertia 
of the star should remain confined in the envelope region in order to have the 
starquake model explanation of Vela glitches. The present study deals with the construction 
of such a model which is possible if we impose the conditions: (i) together with other variables 
(pressure, energy-density, both of the metric parameters $\nu$ and $\lambda$), the (adiabatic) 
speed of sound $v(\equiv \sqrt(dP/dE)$) should also remain continuous at the core-envelope boundary 
of the model governed by an EOS of `adiabatic' polytrope, $dlnP/dln\rho = \Gamma_1$, in the envelope
region and the extreme causal EOS, $(dP/dE) = 1$, in the core region respectively. (ii) The models 
obtained in (i) should follow the `compatibility criterion' that for every assigned value of the 
ratio of central pressure to central energy-density ($\sigma \equiv P_0/E_0$), the compactness ratio $u(\equiv M/R$;
total mass to radius ratio of the static configuration in geometrized units) of the models should remain less than
or equal to the compactness ratio of the corresponding sphere of homogeneous density distribution, 
in order to assure the condition of hydrostatic equilibrium (Negi \& Durgapal 2001; Negi 2004a).
 
The reason for assigning the above mentioned gamma-law EOS for densities below the fiduciary transition density
$E_b$, is not only because of the fact that matching of speed of sound (together with other variables) at the
core-envelope boundary is possible in an analytical manner (Negi \& Durgapal 2000), but also because such models (for various assigned
values of constant $\Gamma_1$) provide the upper bound on NS masses {\em independent} of the EOS of the envelope,
if the condition of `compatibility criterion' is included (Negi 2005). In the present context, however, the value of $\Gamma_1$ 
follows itself from the matching conditions at the core-envelope boundary on the basis of `compatibility criterion'
 and may be looked upon as an `average' (constant) value 
of $\Gamma_1$ below the density range, $E_b$, if this range could have been specified by 
various EOSs like, WFF (Wiringa, Fiks \& Fabrocini 1988), FPS ((Lorenz, Ravenhall \& Pethick, 1993), 
NV (Negele \& Vautherin 1973), or BPS (Baym, Pethick \& Sutherland 1971) in an appropriate sequence, 
as are frequently used by various authors  in the conventional models of NSs (see, e.g. Kalogera \& 
Baym 1996; Friedman \& Ipser 1987), since the choice of the fiduciary transition density, 
$E_b = 1.342 \times 10^{15}\,{\rm g\, cm}^{-3}$ (which is higher than $4E_{nm}$, 
where $E_{nm} =  2.7 \times 10^{14}\,{\rm g\, cm}^{-3}$ denotes the nuclear matter saturation density) 
yields an upper bound on stable NS masses, $M_{max} = 2.2M_\odot$, for our model which, in fact, corresponds to 
the lowest possible upper bound on 
NS masses, independent of $E_b$ if $E_b\geq 4E_{nm}$, calculated by Kalogera \& Baym (1996) on the basis of
modern EOSs for NS matter, fitted to experimental 
nucleon-nucleon scattering data and the properties of light nuclei. This result also
shows consistency with the reasoning mentioned in the previous sentence regarding the 
imposition of the `compatibility criterion' on NS models. The reproduction of the measured 
values of the glitch healing parameter, $Q$, for the Vela pulsar on the basis of the present study 
indicates that this pulsar should be much compact (approximately twice) as compared to that of the Crab 
(see, e.g. Negi 2005) and suggests a further implication (discussed under section 4), if starquake is considered to be
a viable mechanism for glitch generation in all pulsars.

\section{Methodology}
The metric for spherically symmetric and static configurations can be written
in the following form
\begin{equation}
ds^2 =  e^{\nu} dt^2 - e^{\lambda} dr^2 - r^2 d\theta^2 - r^2 $ sin$^2 \theta d\phi^2 , 
\end{equation}
where $\nu$ and $\lambda$ are functions of $r$ alone. The units are assigned in such a manner that the Newtonian
gravitation constant ($G$) and the speed of light in vacuum ($c$) both become unity.  The  Oppenheimer-Volkoff 
(O-V) equations (Oppenheimer \& Volkoff 1939), resulting  from the Einstein's field  equations for the
systems with isotropic pressure $P$  and  energy-density  $E$  can  be 
written as
\begin{eqnarray}
P' & = & - (P + E)[4 \pi P r^3 + m]/r(r - 2m), \\                        
\nu'/2 & = & - P'/(P + E),  \\                                             
m'(r) & = & 4\pi E r^2 \,;
\end{eqnarray}
where the prime denotes radial derivative and $ m(r) $ is the mass contained within the  radius  $r$

\[m(r)  =  \int_{0}^{r} 4\pi Er^2 dr. \] 

The core ($0 \leq r \leq b$) of the present model can be written in the following form
\begin{equation}
 P  =  (E - E_s)
\end{equation}
where  $E_s$  is  the surface density  of the configuration characterized by the extreme causal EOS.
While the envelope ($b \leq r \leq R$) is given by the EOS
\begin{equation}
P  =  K \rho^{\Gamma_1}
\end{equation}
or
                                                
$(E - \rho)  =  P/(\Gamma_1 - 1)$.

\medskip

where $K$ is a constant to be worked out by the matching of  various variables at the 
core-envelope boundary and $\rho$ and $\Gamma_1$ represent respectively, the rest-mass 
density and the (constant) adiabatic index (see, e.g., Tooper 1965).

At the core-envelope boundary, $r = b$, the continuity of $P(=P_b), E(=E_b)$, and 
$r(=r_b)$ require ( Negi \& Durgapal 2000)
\begin{equation}
K = P_b /{(E_b  - [P_b /(\Gamma_1 - 1)])}^{\Gamma_1}
\end{equation}
                                   
where $\Gamma_1$  is given by

\medskip

$\Gamma_1  = [(P + E)/P](dP/dE)$.

\medskip

The continuity of $(dP/dE)$, at the core-envelope boundary requires
\begin{equation}
\Gamma_1  = 1 + (E_b/P_b).
\end{equation}
Thus,  the  continuity  of  $(dP/dE)$, together with other variables ($P, E, \nu$, and $\lambda$) is ensured
at  the core-envelope boundary of the static and spherically 
symmetric configuration.

The coupled Eqs.(3), (4), (5), are  solved  for the model by considering Eq.(6) in the core 
and Eq.(7) in the envelope alongwith the boundary conditions (8) and (9) at 
the  core-envelope  boundary, $ r  =   b $, and   the    boundary  
conditions, $P  = E = 0$ , $m(r = R)  =  M$, $e^{\nu} = e^{-\lambda} = (1 - 2M/R) = 
(1 - 2u)$ at $r  =  R$, at the surface of the configuration, such that for each possible value of
 $\sigma$, the compactness parameter of the whole configuration always turns out to be 
less than or equal to the compactness parameter of the corresponding sphere (with the same $\sigma$) of the
homogeneous density distribution. It is seen that this condition is fulfilled if the {\em minimum} value
 of the ratio of pressure to energy-density, $P_b/E_b$, at the core-envelope boundary reaches 
about $2.92 \times 10^{-1}$. The results of the calculations are presented in Table 1 and 2 respectively, 
while the mass-radius diagram is shown in Fig.1 for a choice\footnote[1]{this choice of transition density
and its subsequent implications are discussed under section 4 of the present study.} 
of $E_b = 1.342 \times 10^{15}\,{\rm g\, cm}^{-3}$ (which is higher than $4E_{nm}$, where $E_{nm} = 2.7 \times 10^{14}\,{\rm g\, cm}^{-3}$ 
is the nuclear matter saturation density). We find that the {\em first} maxima in mass is reached to a value
of $2.2 M_\odot$ for a $\sigma$ value about 0.6285 which is the evidence that the models 
become pulsationally stable up to the maximum value of mass $M_{\rm max} = 2.2 M_\odot$. The radius 
corresponding to this maximum mass is obtained as 9.587 km. Thus, the maximum compactness, $u_{max}$, for 
this stable model yields the value $ \simeq 0.3389$, as shown in Table 1. This upper
bound is found to be fully consistent with the {\em exact} absolute upper bound on compactness ratio of NSs compatible with causality
and pulsationally stability (see, e.g. Negi 2004b), and provides evidence regarding the appropriateness of the model with the $\Gamma_1 = {\rm constant}$ envelope. 
The binding energy per unit rest-mass $\alpha_r [\equiv (M_r - M)/M_r$; where $M_r$ is the rest-mass (see, e.g. Zeldovich \& Novikov 1978)] 
also 
approaches its {\em first} maxima of about 0.2314 for the {\em first} maximum value of mass up to which the configurations remain 
pulsationally stable.


\section{An application of the models to Vela pulsar}

For slowly rotating configurations like Vela pulsar (rotation velocity, $\Omega$, about 70 rad sec$^{-1}$) the moment 
of inertia may be calculated in the first order approximation that appears in the form of Lense-Thirring 
frame dragging-effect. In such situations, however, it appears very useful to use an approximate, but very 
precise empirical formula which is based on the numerical results obtained for thirty theoretical EOSs of dense
nuclear matter. For NSs, the formula yields in the following form (Bejger \& Haensel 2002)
\begin{equation}
I \simeq \frac{2}{9}(1 + 5x)MR^2, \,\,\,\, x > 0.1 
\end{equation}

where $x$ is the compactness parameter measured in units of $[M_\odot({\rm km})/{\rm km}]$, i.e.
\begin{equation}
x = \frac{M/R}{M_\odot/{\rm km}} = \frac{u}{1.477}.
\end{equation}

\begin{figure}
\centering
\psfig{file=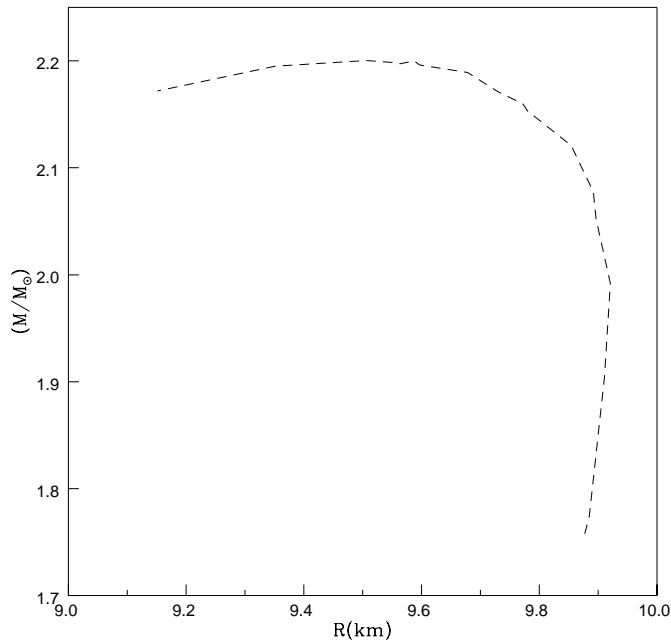,height=9.5cm,width=9.5cm} 
\caption{Mass-Radius diagram of the models as discussed in the text for an 
assigned value of the transition density
$E = E_b = 1.342 \times 10^{15}$ g\, cm$^{-3}$ at the core-envelope
boundary. The  minimum  value  of  the  ratio  of 
pressure to energy-density, $(P_b/E_b)$, at the core envelope boundary is obtained as $2.92 \times 10^{-1}$, such that for an assigned value of $ \sigma $, the inequality $u \leq u_h $  is always satisfied, as shown in Table 1.
              }
\label{FigVibStab}
\end{figure}

 \begin{figure}
   \centering
\psfig{file=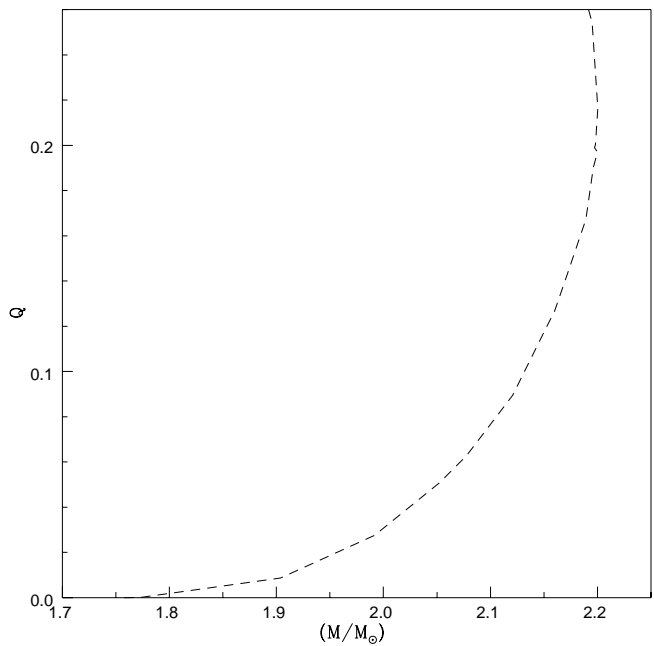,height=9.5cm,width=9.5cm} 
      \caption{Mass ($M/M_\odot$) vs. fractional moment of inertia $Q (=I_{\rm core}/I_{\rm total})$ for the models as discussed in the text and presented by Table 1 and 2 respectively. 
              }
         \label{FigVibStab}
   \end{figure}

Eq.(10) is used, together with coupled Eqs.(3 - 5), to calculate the fractional moment of inertia
given by Eq.(1) and the moment of inertia of the entire configuration for
models presented in Table 1 and Fig.1 respectively. The results are shown in Table 2 and Fig.2 respectively. For the central weighted mean 
value $Q \simeq 0.12$ for the Vela pulsar, Fig.2 yields the mass value $M \simeq 2.153 M_\odot$, radius 
$R \simeq$ 9.780 km, and  surface redshift $z_R \simeq 0.6913$ respectively. The corresponding core mass, 
$M_B$, turns out to be about $0.810M_\odot$ as shown in Table 2 which is about 37.6$\%$ of the total mass 
of the entire configuration. This value of Vela mass can exceed upto a value about $2.196 M_\odot$ if the 
upper limit of the central weighted mean value of $Q \simeq 0.19$ is considered. The corresponding values 
of surface redshift and core mass can exceed upto the values 0.7568 and $1.046 M_\odot$ respectively. This 
value of surface redshift ($z_R \simeq 0.7568$), in fact, represents an ultra-compact object (UCO; $z_R \geq 
0.73$) which are entities of interest (see, e.g. Negi \& Durgapal 1999; and references therein). 
For the lower limit of the `central' weighted mean value of $Q \simeq 0.05$, the Vela mass can reduce upto a 
value of about $2.052 M_\odot$. 
The corresponding values of surface redshift and core mass turn out to be 0.6066 and $0.489M_\odot$ 
respectively. Thus, the core mass of the structure varies between 23.8 $\%$ to 47.6$\%$ of the total 
mass, for the lower limit of $Q \simeq 0.05$ to the upper limit of $Q \simeq 0.19$ respectively. For $Q$ 
values larger than about 0.197 (i.e., the point of maximum mass $M_{max} = 2.2 M_\odot$), the structures 
become pulsationally unstable, whereas  the minimum stable Vela mass corresponds to a value about 1.758 
$M_\odot$ as $Q \rightarrow 0$.


\section{Discussion}
In the previous study, we considered the values of constant $\Gamma_1$ = (4/3), (5/3), and
2 respectively for the density range below the fiduciary transition density $E_b$ on the basis of `compatibility
criterion' in order to construct the starquake models for Crab pulsar (Negi 2005). If (i) the observational
constraint of the glitch healing parameter, and (ii) the observational constraint of the 
recently evaluated value of the moment of inertia 
  for the Crab pulsar were combined together with the `compatibility criterion' mentioned 
  above, the model with $\Gamma_1 = (5/3)$
  envelope itself yielded  the value of transitions density, $ \simeq 2.7 \times 
  10^{14}\,{\rm g\, cm}^{-3}$, the nuclear matter saturation density at the core-envelope boundary.
  This value of $E_b $ yields 
  the upper bound on NS masses $M_{\rm max} \simeq 4.1 M_\odot$, {\em independent} of the 
  EOS of the envelope (Negi 2005). Not only this value is found 
  fully consistent with the upper bound on NS masses obtained by using modern EOSs for NS 
  matter, fitted to experimental 
nucleon-nucleon scattering data and the properties of light nuclei, consistent with causality 
and dynamical stability (see, Fig.2 of Kalogera \& Baym 1996, for a fiduciary transition 
density $E_b = 2.7 \times 10^{14}\,{\rm g\, cm}^{-3}$), the upper bound on compactness 
ratio of these models  ($u_{max} \simeq 0.34$) was also found fully consistent with the
`absolute' upper bound on compactness 
ratio of NSs consistent with causality and pulsational stability (Negi 2004b). The results of 
this study are also supported by observations, since the existed value of the upper bound on 
the energy of a gravitationally redshifted $\gamma$-ray line at about 0.40 MeV from the Crab 
pulsar is found consistent with the predicted value in the energy range of about 0.414 - 
418 MeV (see, Negi 2005; and references therein).

In the previous study, the  value of the ratio of pressure to 
density at the core-envelope boundary, $P_b/E_b$, on the basis of `compatibility criterion'
was calculated as $\sim 1.065 \times 10^{-2}$, which represents the region very close to the surface
of the star (thin crust), and does not require the necessary continuity of the speed of sound (together
with other variables) at this boundary which (obviously) corresponds to a region of relatively low density.
Since, we were concerned with the larger values of the glitch healing parameter, $Q \geq 0.70$, for the
Crab pulsar which demanded that more than 70$\%$ of the moment of inertia should remain attached to 
the core component. However, the much smaller values of the glitch healing parameter for Vela pulsar, 
$Q \leq 0.20$, require that more than 80$\%$ of the moment of inertia should remain confined to 
the envelope region which obviously require a much larger value of $P_b/E_b$ (thick crust) as compared to the previous case, i.e.
a much deeper region of higher density inside 
the crust where one can naturally assume that the speed of sound approaches the 
speed of light and is continuous at some `fiduciary' density at the boundary of extreme causal 
EOS in the core. In the present context of the Vela pulsar, we do not have any information 
about the Vela's moment of inertia, as we did have in the case for the Crab. However, if the Vela and the Crab
both are members of the same conventional sequence of the NS models which terminates at the maximum value of
NS mass, then we can assign the value of $4.1 M_\odot$ as an 
upper bound of NS masses obtained in the previous study, in order to constrain the present models. This assignment
gives the Vela's mass about  $4.01 M_\odot$ for the `central' weighted mean value 
of the glitch healing parameter $Q \simeq 0.12$ which looks too large, since none of the observational studies claim such higher values of mass for the Vela pulsar. The 
corresponding value of transition density, $E_b$, turns out to be about $3.864 \times 10^{14}\,
{\rm g\, cm}^{-3}$, which also seems unlikely for the normalization of the $P_b/E_b$ ratio of $\sim 2.92 \times 
10^{-1}$ calculated for the present models. Notwithstanding, the result of the previous study 
which is found fully consistent with the present one is the upper bound on NS compactness, 
$u \simeq 0.34$.

This finding reveals that if starquake 
 is considered to be a viable model for glitch generation in the Vela pulsar\footnote[2]{The 
 other problems associated with the starquake model for Vela pulsar, like the large size of glitches 
 ($\Delta\Omega/\Omega \sim 10^{-6}$), and the observations of significant amount of change in X-ray 
 flux soon after the glitch occurs (see, e.g. Crawford \& Demia\'{n}ski (2003); and references therein
 ) are not discussed in the present paper. However, in view of the higher values of compactness ratio of the present models,
 the future study in this regard may provide some explanation.} 
 together with the Crab (and, in fact, in all pulsars!) then it would lead to the
 following implication that there might exist a parallel sequence of the stable NS masses composed of
 Vela-like pulsars, 
 together with the conventional sequence which composes the Crab-like pulsars, such
 that this parallel sequence begins with
somewhat higher value of the compactness ratio as compared to the conventional sequence
but terminates with almost the same  value of maximum compactness (e.g., the parallel sequence considered
in the present study begins with a value of compactness ratio about 0.2628 which is higher
than the corresponding value of compactness ratio of conventional sequence
undertaken by Negi (2005), but both of these sequences terminate with almost the same value of 
maximum compactness, $u \simeq 0.34$) consistent with
 the absolute upper bound on NS 
 compactness (Negi 2004b). However, such a sequence would correspond to different
 values of minimum  and maximum 
  stable masses as compared to the conventional sequence. Apparently, the minimum mass
  of this sequence would be higher than the minimum mass of the conventional sequence,
  whereas the maximum mass of this sequence would correspond to a value less than the
  maximum mass of the conventional sequence.

Kalogera \& Baym (1996) have shown that the lowest possible upper bound on NS masses corresponds to a value of $2.2 M_\odot$ independent of $E_b$, provided the
 fiduciary transition density, $E_b$, satisfies the condition $E_b \geq 4E_{nm}$. Our models, in fact,
 satisfy this condition, since for the lowest possible upper bound of $2.2 M_\odot$, we obtain the value of
transition density, $E_b = 1.342 
\times 10^{15}\,{\rm g\, cm}^{-3}$, which is higher than $ 4E_{nm}$.  
 Apparently, this value of transition density yields the `minimum' 
possible mass for Vela pulsar on the basis of present models as $2.153 M_\odot
$ for the `central' weighted mean 
value of $Q \simeq 0.12$. This is discussed earlier under sections 2 and 3 respectively.
Thus, in view of the maximum mass and density regime, the results of the present study are fully 
consistent with the model based on the modern theory of nuclear 
interactions and underline the appropriateness of the use of gamma-law EOS in the envelope, the necessity
of the continuity of speed of sound across the boundary, and the imposition of 
the `compatibility criterion', in order to have a starquake model explanation for Vela-like pulsars.

\section{Results and conclusions}

The study shows that if starquake is responsible for the glitch generation in Vela pulsar, the 
surface redshift, $z_R$, for NS model of Vela pulsar should correspond to a value of 0.6912, if the 
observational constraint of the `central' weighted mean value of the glitch healing parameter $Q \simeq 
0.12$ is imposed. 
This value  gives the `minimum' mass  of the Vela 
 pulsar $M \simeq 2.153 M_\odot$ for the `transition density' $E_b = 1.342 \times 10^{15}{\rm\,g\,cm}^{-3}$
 at the core-envelope 
 boundary, if the lowest possible upper bound on NS masses 
 is considered to be $2.2 M_\odot$ . The mass is slightly increased upto the value $M \simeq 2.196 
 M_\odot$, but the surface redshift can increase upto the value $z_R \simeq 0.7568 $ if the 
 observational constraint of the upper weighted mean value of the glitch healing parameter $Q 
 \simeq 0.19$ is imposed. However, for the lower weighted mean value of the glitch healing parameter 
 $Q \simeq 0.05$, the mass and surface redshift can decrease upto $z_R \simeq 0.6066$ and $M \simeq 
 2.052M_\odot$ respectively. For  the upper weighted mean value of  $Q \simeq 0.19$, the structures 
 represent ultra-compact objects (UCO) which are entities of important astrophysical interest 
 (see, e.g. Negi \& Durgapal 1999; and references therein). In the limiting case of $Q\rightarrow 0$, the minimum
 mass for Vela corresponds to the value of $1.758M_\odot$.
 The confirmation of these results require an observation of the lower bound on the energy of a 
 gravitationally redshifted $\gamma$-ray line in the narrow energy range about 0.291 - 0.302 MeV 
 from the Vela pulsar which is in 
 excellent agreement with the observation of the lower bound on the energy
 of a $\gamma$-ray pulse of 0.30 MeV from the Vela pulsar (Tumer et al. 1984), if this energy could be 
 interpreted as the 
 energy of a gravitationally redshifted electron-positron annihilation radiation from the 
 star's surface.
 
\section*{Acknowledgments}

The author acknowledges the Aryabhatta Research Institute of Observational Sciences (ARIES), Nainital 
for providing library and computer-centre facilities.

\begin{table*}

\begin{center}
      \caption[]{Various values of mass ($M/M_\odot$), radius $R({\rm km})$, compactness ratio ($u$), binding-energy per particle ($\alpha_r$), corresponding to the models discussed in the text for  different assigned values  of $ \sigma $. The compactness ratio for  homogeneous  
density distribution is represented by $u_h$. The slanted values  correspond  to 
the limiting case upto which the configuration remains pulsationally stable.}

\begin{tabular}{cccccc}

\hline

${(P_0 / E_0)}$ & $M/M_\odot$  & $R({\rm km})$  &  $\alpha_r$  & $u$ & $u_h$  \\
 
\hline
0.29202 & 1.757659 & 9.877204 & 0.169796 & 0.262834 & 0.262854 \\
0.29800 & 1.772991 & 9.884901 & 0.171393 & 0.264920 & 0.265166 \\  
0.35452 & 1.903785 & 9.910778 & 0.186960 & 0.283720 & 0.284570 \\  
0.39993 & 1.991678 & 9.920844 & 0.198449 & 0.296518 & 0.297502 \\
0.43923 & 2.052262 & 9.896808 & 0.206909 & 0.306280 & 0.307194 \\ 
0.45699 & 2.077665 & 9.892000 & 0.210533 & 0.310222 & 0.311190 \\ 
0.49397 & 2.120681 & 9.854696 & 0.217014 & 0.317843 & 0.318832 \\  
0.53331 & 2.153382 & 9.780031 & 0.222426 & 0.325208 & 0.326097 \\  
0.54099 & 2.159962 & 9.772185 & 0.223498 & 0.326464 & 0.327424 \\   
0.55999 & 2.171257 & 9.729357 & 0.225572 & 0.329615 & 0.330584 \\  
0.59146 & 2.189320 & 9.678408 & 0.228886 & 0.334107 & 0.335476 \\ 
0.61902 & 2.196170 & 9.597157 & 0.230666 & 0.337990 & 0.339441 \\  
{\sl 0.62850} & {\sl 2.200000} & {\sl 9.587457} & {\sl 0.231448} & {\sl 0.338922} & 0.340741 \\  
0.62936 & 2.197304 & 9.563581 & 0.231138 & 0.339352 & 0.340858 \\  
0.63148 & 2.198152 & 9.560925 & 0.231311 & 0.339577 & 0.341144 \\
0.65148 & 2.200457 & 9.506329 & 0.232207 & 0.341885 & 0.343769 \\
0.69998 & 2.195136 & 9.351309 & 0.232781 & 0.346713 & 0.349633 \\
0.75294 & 2.171828 & 9.151000 & 0.230592 & 0.350540 & 0.355328 \\

\hline

\end{tabular}

\end{center}
\end{table*}


\begin{table*}

\begin{center}
      \caption[]{Various values of mass ($M/M_\odot$), radius $R({\rm km})$, core mass ($M_b/M_\odot$), core radius  $R_b({\rm km})$, and fractional moment of inertia $I_{\rm core}/I_{\rm total}$, corresponding to the models discussed in the text for  different assigned values  of $ \sigma $. The corresponding surface and central redshifts are represented by  $z_R$ and $z_0$ respectively. The slanted values  correspond  to 
the limiting case upto which the configuration remains pulsationally stable.}

\begin{tabular}{cccccccc}

\hline

${(P_0 / E_0)}$ & $M/M_\odot$  & $R({\rm km})$  &  $M_b/M_\odot$  & $R_b({\rm km})$ & $I_{\rm core}/I_{\rm total}$ & $z_R$ & $z_0$ \\
 
\hline
0.29202 & 1.757659 & 9.877204 & 0.000001 & 0.084931 & 0.000000 & 0.451973 & 0.961475 \\
0.29800 & 1.772991 & 9.884901 & 0.006151 & 1.301636 & 0.000032 & 0.458402 & 0.983101 \\
0.35452 & 1.903785 & 9.910778 & 0.171574 & 3.889230 & 0.008641 & 0.520468 & 1.202564 \\  
0.39993 & 1.991678 & 9.920844 & 0.341592 & 4.844349 & 0.027604 & 0.567552 & 1.394268 \\
0.43923 & 2.052262 & 9.896808 & 0.489306 & 5.412447 & 0.050835 & 0.606561 & 1.573780 \\
0.45699 & 2.077665 & 9.892000 & 0.554063 & 5.617970 & 0.062644 & 0.623161 & 1.658789 \\
0.49397 & 2.120681 & 9.854696 & 0.683071 & 5.969921 & 0.089515 & 0.656770 & 1.846732 \\
0.53331 & 2.153382 & 9.780031 & 0.809999 & 6.255151 & 0.120661 & 0.691314 & 2.065669 \\
0.54099 & 2.159962 & 9.772185 & 0.833445 & 6.301950 & 0.126634 & 0.697422 & 2.110140 \\
0.55999 & 2.171257 & 9.729357 & 0.889491 & 6.406664 & 0.142236 & 0.713049 & 2.225541 \\
0.59146 & 2.189320 & 9.678408 & 0.976141 & 6.548950 & 0.167190 & 0.736086 & 2.426520 \\ 
0.61902 & 2.196170 & 9.597157 & 1.045560 & 6.644820 & 0.190181 & 0.756766 & 2.621490 \\ 
{\sl 0.62850} & {\sl 2.200000} & {\sl 9.587457} & {\sl 1.068060} & {\sl 6.672101} & {\sl 0.197133} & {\sl 0.761842} & {\sl 2.688782} \\
0.62936 & 2.197304 & 9.563581 & 1.070062 & 6.674508 & 0.198876 & 0.764196 & 2.698936 \\
0.63148 & 2.198152 & 9.560925 & 1.074988 & 6.680161 & 0.200427 & 0.765435 & 2.714632 \\
0.65148 & 2.200457 & 9.506329 & 1.119576 & 6.727057 & 0.216372 & 0.778276 & 2.871016 \\
0.69998 & 2.195136 & 9.351309 & 1.214118 & 6.793253 & 0.254274 & 0.806058 & 3.299069 \\
0.75294 & 2.171828 & 9.151000  &1.294603 & 6.792263 & 0.293308 & 0.829036 & 3.871991 \\

\hline

\end{tabular}

\end{center}
\end{table*}


\vspace{1.0cm}

\end{document}